\documentclass[10pt,twocolumn]{article}
\usepackage[margin=1in]{geometry}
\usepackage{amsmath}
\usepackage{abstract}
\renewcommand{\thesection}{\Roman{section}}
\usepackage{amsfonts,color}
\usepackage{amssymb}
\usepackage{graphicx}
\usepackage{hyperref}
\begin{document}
\title{\bf{Magnetic Field of a compact spherical star under $f(R,T)$ gravity}}
\author
{Safiqul Islam$^\ast$ and Shantanu Basu$^\dagger$\\
$^\ast$Harish-Chandra research Institute, HBNI, Chhatnag Road, Jhunsi, Allahabad-211019, India \\
$^\dagger$Department of Physics and Astronomy, The University of Western Ontario,\\ 1151 Richmond Street, London, ON, Canada, N6A 3K7\\
$^\ast$safiqulislam@hri.res.in \\ $^\dagger$basu@uwo.ca
}
\twocolumn[
\maketitle
\begin{onecolabstract}
We present the interior solutions of distributions of magnetised fluid inside a sphere in
$f(R,T)$ gravity. The magnetised sphere is embedded in an exterior Reissner-Nordstr\"{o}m metric. We assume that all physical quantities are in static equilibrium. The perfect fluid matter is studied under a particular form of the Lagrangian density $f(R,T)$. The magnetic field profile in modified gravity is calculated. Observational data of neutron stars are used to plot suitable models of magnetised compact objects. We reveal the effect of $f(R,T)$ gravity on the magnetic field profile, with application to neutron stars, especially highly magnetized neutron stars found in X-ray pulsar systems. Finally the effective potential $V_{\rm eff}$ and innermost stable circular orbits, arising out of motion of a test particle of negligible mass influenced by attraction or repulsion from the massive center, are discussed.
\end{onecolabstract}
]


\section{Introduction}

~~~In a homogeneous and isotropic spacetime in General Relativity (GR), the Friedmann equations
that describe the evolution of the universe are a result of the Einstein field
equations. Observational cosmology has revealed that the universe has undergone two phases of cosmic acceleration.
The first one is called inflation era
which is believed to be the precursor of the radiation dominated era \cite{u}. This phase is a necessity
not only to solve the flatness and horizon
problems related to big-bang cosmology, but also to explain the intrinsic details of
a nearly flat spectrum of temperature disparities
observed in Cosmic Microwave Background (CMB) \cite{s}. It was only after the matter dominated
era that the second accelerating phase started. The unknown component giving rise to this late
time cosmic acceleration is called dark energy \cite{h}.
The existence of dark energy has been confirmed by a number of observations, such as
supernovae Ia (SN Ia) \cite{q},
large-scale structure (LSS) \cite{v}, baryon acoustic oscillations
(BAO) \cite{e} and CMB \cite{t}.

Though the modifications of the energy-momentum tensor in Einstein equations
give rise to the scalar-field models of inflation and dark energy, there is yet
another approach to explain the acceleration of the universe. This relevant field is the modified gravity in which the gravitational theory is
modified in comparison to GR. The Lagrangian density for GR is given by $f(R)=R-2 {\Lambda}$, where $R$ is
the Ricci scalar and $\Lambda$ refers to the cosmological constant (corresponding to the equation of state $w=-1$).
The presence of $\Lambda$ gives rise to an exponential expansion of the universe, but it is not possible to use it for
inflation because the inflationary period needs to be connected to the radiation era. It is however possible to use the cosmological constant to represent dark energy and use it to model the accelerating expansion of the present-day universe. In contrast however, if the cosmological constant originates from a vacuum energy of particle physics, its energy density would exceed the dark energy density of the present universe. The $\Lambda$-Cold Dark Matter ($\Lambda$CDM) model $(f(R)=R-2 \Lambda)$ satisfies a number of observational data \cite{i}, but there is also a possibility of a time-varying equation of state of dark energy \cite{n}.

Recently, a relativistically covariant model of interacting dark energy based on the principle of least action has been proposed \cite{o}. They observed that the cosmological term $\Lambda$ in the gravitational Lagrangian is dependent on the trace of the energy-momentum tensor $T$ and consequently the model was termed as $\Lambda(T)$ gravity. Subsequently another extension of standard GR was considered \cite{f}. These are the $f(R,T)$ modified theories of gravity. The energy conditions and thermodynamics in $f(R,T)$ theories have been investigated \cite{r}. Stationary electromagnetic fields of slowly rotating relativistic magnetized
star in the braneworld has been studied \cite{rrr}.

There are two formalisms in deriving field equations from the action in $f(R,T)$ gravity. The first is the standard metric formalism in which the field equations are derived by the variation of the action with respect to the metric tensor $g_{\mu \nu}$.
In this formalism the affine connection ${\varGamma}_{\beta \gamma}^{\alpha}$
depends on $g_{\mu \nu}$. In this paper we have followed a similar approach. The second is the Palatini formalism in which $g_{\mu \nu}$ and ${\varGamma}_{\beta \gamma}^{\alpha}$
are treated as independent variables when we vary the action.

The organization of this paper is envisaged as follows:
The field equations under $f(R,T)$ gravity are briefed in section II. In section III we study the magnetic field inside the star, with the imposition of vanishing electric current density in its exterior region $[J^{r} = J^{\theta} = J^{\phi} = 0]$. The magnetic field outside the star is also calculated. Section IV deals with the test motion of electrically charged particles in this space time geometry. Concluding remarks are in section V.

\section{Field equations in the metric formalism of $f(R,T)$ gravity}

~~~We start with the 4-dimensional action in $f(R,T)$ gravity \cite{f} as
\begin{equation}
S=\frac{1}{16\pi}\int d^{4}x f(R,T) \sqrt{-g} + \int d^{4}x {\mathcal{L} _ {m}}\sqrt{-g} .
\end{equation}
Here $g$ is the determinant of the metric tensor $g_{\mu \nu}$, $\mathcal{L} _ {m}$ is a matter Lagrangian, $R$ is the Ricci scalar, $T$ is the trace of the stress-energy tensor of the matter $T_{\mu\nu}$ \cite{j}, and $f(R,T)$ is the Lagrangian density in GR.
Here different choices of $\mathcal{L} _ {m}$ can be
considered, each of which directs to a specific form of fluid. 
We assume the geometric units $c=G=1$.

On varying the above action of eqn. (1) with respect to the metric $g_{\mu\nu}$, the following field equations of $f(R,T)$ gravity are evident:

\begin{eqnarray}
f_{R}(R,T)R_{\mu\nu}-\frac{1}{2}f(R,T)g_{\mu\nu}\nonumber\\
+(g_{\mu\nu}\square-\triangledown_{\mu}\triangledown_{\nu})f_{R}(R,T)\nonumber\\
= 8{\pi}T-f_{T}(R,T)T_{\mu\nu}-f_{T}(R,T)\Theta_{\mu\nu},
\end{eqnarray}
where
\begin{eqnarray}
f_{R}(R,T)=\frac{\partial f(R,T)}{dR}, f_{T}(R,T)=\frac{\partial f(R,T)}{dT}, \nonumber\\
{\square} {\equiv} {\partial}_{\mu} ({\sqrt{-g} g^{\mu \nu} \partial_{\nu}})/{\sqrt{-g}}.
\end{eqnarray}

The covariant divergence of the stress energy tensor \cite{z} yields
\begin{eqnarray}
\triangledown^{\mu}T_{\mu \nu}&=&\frac{f_{T}(R,T)}{8 \pi -f_{T}(R,T)}[(T_{\mu \nu}+\Theta_{\mu\nu}) \triangledown^{\mu}ln f_{T}(R,T) , \nonumber\\
& &+ \triangledown^{\mu}\Theta_{\mu\nu}-\frac{1}{2} g_{\mu\nu} \triangledown^{\mu}T.
\end{eqnarray}

In this paper we consider the energy-momentum tensor of a perfect fluid type as
\begin{equation}
T_{\mu\nu}=(\rho+p)u_{\mu}u_{\nu}-p g_{\mu\nu},
\end{equation}

where the space-like four velocity vector $u^{\mu}u_{\mu}=1$ and $u^{\mu}\triangledown_{\nu} u_{\mu}=0$.

Next, we take the functional form $f(R,T)=R+ 2 f(T)= R+ 2 {\kappa}^{n} T$, where $f(T)={\kappa}^n T$ for which
the conditions imposed are

\begin{equation}
\mathcal{L} _ {m}=-p, \Theta = -2T_{\mu\nu}-pg_{\mu\nu}.
\end{equation}

Hence the Einstein eqn. in GR reduces to [\cite{l},\cite{m},\cite{d}],
\begin{equation}
G_{\mu\nu}=  8 \pi T_{\mu\nu}+ {\kappa}^n T g_{\mu\nu}+ 2 {\kappa}^n (T_{\mu\nu} + p g_{\mu\nu}) .
\end{equation}

The energy momentum tensors are conserved as in GR in eqn. (7) on equating $\kappa=0$, i.e., when $f(R,T)=R$.

\section{Magnetic field of the star within the $f(R,T)$ gravity}

~~~The line element for spherically symmetric metric describing the charged
compact star stellar configuration is given in the Reissner-Nordstr\"{o}m (RN) metric by

\begin{eqnarray}
ds^2 &=& -(1-\frac{2m(r)}{r}+\frac{q^2(r)}{r^2}) dt^{2}\nonumber\\
& &+(1-\frac{2m(r)}{r}+\frac{q^2(r)}{r^2})^{-1} dr^{2} \nonumber\\
& &+r^2(d\theta^{2}+\sin^{2}\theta d\phi^{2}),
\end{eqnarray}
where $m(r)$ and $q(r)$ are the mass function and electric charge function of the perfect fluid, respectively .

The magnetic field of the star within the $f(R,T)$ gravity” is based on solutions of the Maxwell equations for the test magnetic field assuming that the magnetic field configuration of the star is dipolar. We assume that the components of the magnetic field $B(r, \theta)$ inside the star \cite{aa} are

\begin{eqnarray}
B^{\hat{r}}(r,\theta) = F(r) \cos {\theta} ,\nonumber\\
B^{\hat{\theta}}(r,\theta) = G(r) \sin {\theta} ,\nonumber\\
B^{\hat{\phi}}(r,\theta) = 0 .
\end{eqnarray}

Here the unknown radial functions $F(r)$ and $G(r)$ are responsible for the relativistic corrections in the gravitational field of $f(R,T)$ gravity. In the exterior region, we impose the condition of vanishing electric current density, i.e., $J^{\hat{r}} = J^{\hat{\theta}} = J^{\hat{\phi}} = 0$. We find the Maxwell equations for the radial part of the magnetic field [\cite{bb}, \cite{ccc}] to be

\begin{eqnarray}
(r^2 F(r))_{,r}+ 2 r G(r) \sqrt{D(r)} = 0 , \nonumber\\
(r G(r)\sqrt{A(r)})_{,r}+F(r) \sqrt{A(r)D(r)} = 0,
\end{eqnarray}
where the analytic expressions for the metric functions $A(r)$ and $D(r)$ exterior to the star are defined [\cite{aa},\cite{cc}-\cite{gg}] as

\begin{equation}
N^2 (r) = A(r) = D^{-1} = (1-\frac{2m(r)}{r}+\frac{q^2(r)}{r^2}).
\end{equation}

The unknown radial function $F(r)$ is obtained from a second-order ordinary differential equation [\cite{aa},\cite{ff}]:

\begin{eqnarray}
\frac{d}{dr}[(1-\frac{2m(r)}{r}+\frac{q^2(r)}{r^2}) \sqrt{2}\frac{d}{dr}(r^2 F(r))\nonumber\\-\sqrt{2} F(r)] = 0.
\end{eqnarray}

On simplification the above eqn. (12) reduces to

\begin{eqnarray}
F''(r)[r^2 - 2 r m(r)+q^2(r)]+F'(r)[4 r \nonumber\\
-6m(r)-2 r m'(r)+\frac{2}{r}+ 2 q(r) q'(r)]\nonumber\\
+F(r)[1-4m'(r)-\frac{2}{r^2}q^2(r)
+ \frac{4}{r}q(r) q'(r)]\nonumber\\
=0 .
\end{eqnarray}
However as $r\rightarrow \infty$, $\frac{m(r)}{r}\rightarrow 0$ and also $\frac{q^2(r)}{r^2}\rightarrow 0$ and they do not contribute to the magnetic field [\cite{hh}-\cite{lll}]. Hence eqn. (13) reduces to
\begin{equation}
r^2 F''(r) + 4 r F'(r) +  F(r) = 0,
\end{equation}
which yields
\begin{equation}
F(r) = a_1 r^{m_1} + b_1 r^{n_1},
\end{equation}
where $a_1$ and $b_1$ are arbitrary constants and $m_1=-\frac{3}{2}-\frac{\sqrt{5}}{2}$, $n_1=-\frac{3}{2}+\frac{\sqrt{5}}{2}$.\\

Using eqns. (10) and (15) we find that
\begin{equation}
G(r) = \frac{1}{2}(-2 a_1 r^{m_1} - a_1 m_1 r^{m_1} - 2 b_1 r^{n_1}  - b_1 n_1 r^{n_1}).
\end{equation}

At the poles $\theta=0$, and the magnetic field from eqn. (9) depends only on $F(r)$. On the other hand when we consider the equatorial plane, $\theta=\frac{\pi}{2}$, $B(r,\theta)$ then depends only on $G(r)$, which gives the magnetic field in $f(R,T)$ gravity.\\

The dipolar magnetic field in GR is derived [\cite{aa},\cite{ccc},\cite{mm}] to be
\begin{equation}
\frac{G_{\rm GR}(r)}{G_{\rm Newt}(r)} = -\frac{3 R_1^3}{8 M^3}\left[ \ln N^2 + \frac{2 M}{r}(1+ \frac{M}{r})\right],
\end{equation}
where $M$ and $R_1$ stand for the mass and radius of the star, respectively, and the magnetic field at the pole in
the Newtonian limit is
\begin{equation}
G_{\rm Newt}(r) = \frac{2 \mu}{r^3},
\end{equation}
with $\mu$ being constant.\\

The radial dependence of the magnetic field in eqn. (16) for various values of constants ($a_1,b_1$)) is plotted in Fig. 1.

\begin{figure}[htbp]
	\centering
	\includegraphics[scale=.85]{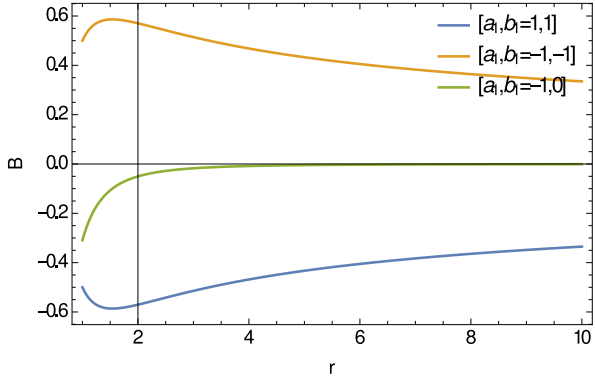}
	\caption{The magnetic field $B ({\rm km}^{-3})$ in $f(R,T)$ modified gravity for various values of constants $(a_1,b_1)$ is shown versus radius $r ({\rm km})$.}
\end{figure} 

It is obvious from eqn. (11) that as $r\rightarrow \infty$, $\ln N^2\rightarrow 0$
Hence eqn. (17) reduces to,

\begin{equation}
G_{\rm GR}(r) = -\frac{3  \mu  R_1^3}{2 M^2 r^4}(1+ \frac{M}{r}).
\end{equation}

We consider below four different models ($M_{\odot}$ being the solar mass) using stellar mass data [\cite{nn},\cite{oo}]:
as (i) X-ray pulsar, Her X-1 [\cite{nn},\cite{oo},\cite{pp},\cite{qq}], which is characterized by mass
$M=1.47 M_{\odot}$ and radius $R_1=4.921$ km.
(ii) X-ray pulsar, 4U 1700-37 [\cite{nn},\cite{oo}], which is characterized by mass
$M=2.44 M_{\odot}$ and radius $R_1=8.197$ km.
(iii) Neutron star, J1518+4904 [\cite{nn},\cite{oo}], which is characterized by mass
$M=0.72 M_{\odot}$ and radius $R_1=2.419$ km.
(iv) Neutron star, J1748-2021 B [\cite{nn},\cite{oo}], which is characterized by mass
$M=2.74 M_{\odot}$ and radius $R_1=9.281$ km.\\

We also plot the magnetic field $B$ using the general relativistic value given by eqn. (19) outside the star in Fig. 2. and Fig. 3
\begin{figure}[htbp]
	\centering
	\includegraphics[scale=0.9]{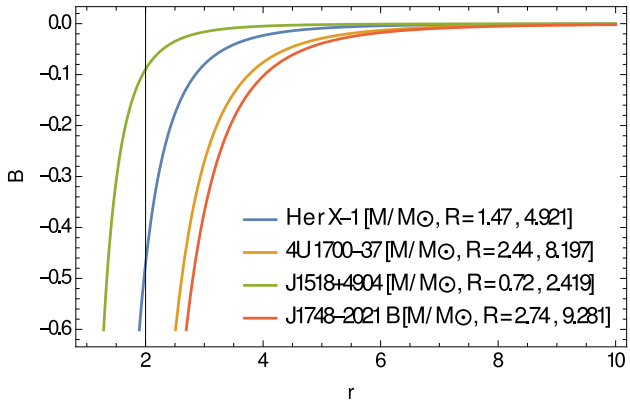}
	\caption{The magnetic field $B ({\rm km}^{-3})$ in GR $(\mu=1/28)$ is shown versus $r ({\rm km})$ for various observed objects.}
\end{figure}

\begin{figure}[htbp]
	\centering
	\includegraphics[scale=0.9]{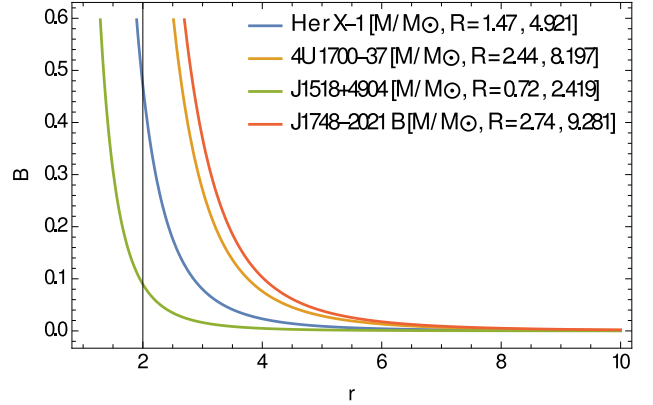}
	\caption{The magnetic field $B ({\rm km}^{-3})$ in GR $(\mu=-1/28)$ is shown versus $r ({\rm km})$ for the same objects as in Fig. 2.}
\end{figure}

Here the constant $\mu$ can have any value, but we have chosen it as, $\mu=-1/28$ and $\mu=1/28$ for convenience.

For a particular choice of $\mu=-1/28$, Fig. 4 and Fig. 5 further specify the magnetic field of X-ray pulsar, Her X-1, both in the interior and exterior regions. 

Here we consider the geometric units with $c = G = 1$.

\begin{figure}[htbp]
	\centering
	\includegraphics[scale=0.85]{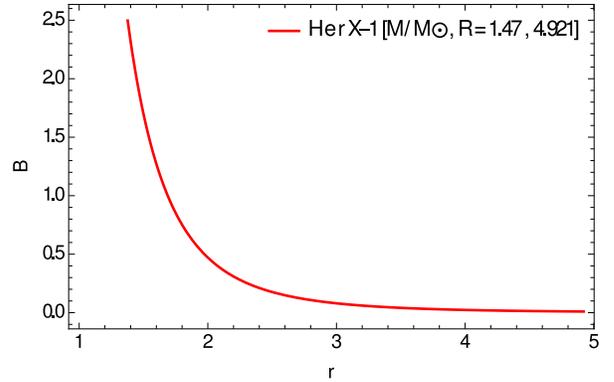}
	\caption{The magnetic field $B ({\rm km}^{-3})$ in GR of X-ray pulsar, Her X-1 in the interior region is shown versus $r ({\rm km})$.}
\end{figure}
\begin{figure}[htbp]
	\centering
	\includegraphics[scale=0.85]{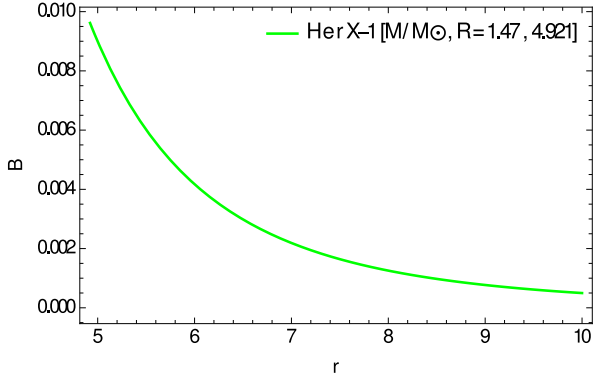}
	\caption{The magnetic field $B ({\rm km}^{-3})$ in GR of X-ray pulsar, Her X-1 in the exterior region is shown versus $r ({\rm km})$.}
\end{figure}

\section{Effective potential}

~~~We study the effective potential $V_{\rm eff}$, as a result of motion of a test particle of negligible mass undergoing attraction or repulsion by a massive static center. So, first we deduce the non-vanishing components of the electromagnetic field four-potential $A_{\mu}$ as [\cite{ssss},\cite{sssss}], considering $B_0=B^z_{0}=B^{\hat{\theta}}(r,\theta)$ as the component aligned along the axis; $B_1=B^x_{0}=B^{\hat{r}}(r,\theta)$ as the component perpendicular to the rotation axes; and $B^y_{0}=B^{\hat{\phi}}(r,\theta)$. Here $B^x_{0}, B^y_{0}, B^z_{0}$ denote the components of the field in asymptotically Minkowskian coordinates [$x=r \sin \theta \cos \phi, y=r \sin \theta \sin \phi, z=r \cos \theta$]. The electric charges being enclosed in 3-volume $\Sigma$, $A_{\mu}$ is derived from the electromagnetic field tensor $F_{{\mu}{\nu}}$=$A_{\nu, \mu}-A_{\mu, \nu}$ [\cite{ssss},\cite{sss}], using eqn. (9), as
\begin{eqnarray}
A_{t}=-\frac{Qr}{\Sigma}\,;~~~A_{r}=0\,;~A_{\theta}=0\,;\nonumber\\
A_{\phi}=\frac{1}{2} B_{0} r^2=\frac{r^2}{2} G(r) \sin \theta.
\end{eqnarray}
~~With the variables and parameters expressed by the scale transformations $t\to tM, r\to rM, B_{0}\to B_{0}/M, \tau \to \tau M, Q \to QM$, where $m(r)\to M$, $q(r)\to Q$ being the mass and charge of the black hole and $\tau$ is the proper time, we obtain the dimensionless Lagrangian using eqn. (8), with $ds^2=-d\tau^2$, as
\begin{eqnarray}
\mathcal{L}=\frac{1}{2}[-(1-\frac{2}{r}+\frac{Q^2}{r^2})\dot{t}^2+(1-\frac{2}{r}+\frac{Q^2}{r^2})^{-1}\dot{r}^2\nonumber\\
+r^2 \dot{\theta}^2+r^2 \sin^2 \theta \dot{\phi}^2]. 
\end{eqnarray}
The Lagrangian $\mathcal{L}$ is identical to $-{1/2}$. The motion of electrically charged particles are described by the Hamiltonian \cite{qqq},
\begin{eqnarray}
\mathbb{H}=\frac{1}{2}g^{\alpha \beta}(p_\alpha-qA_{\alpha})(p_\beta-qA_{\beta}),
\end{eqnarray}
where $P_{\alpha}=p_\alpha-qA_{\alpha}=\frac{\partial \mathcal{L}}{\partial \dot{x}^{\beta}}=g_{\alpha \beta}\dot{x}^{\beta}$, $\alpha, \beta \in$ $\left\lbrace 0,1,2,3\right\rbrace$ and $q$ is the electric charge of the test particle with the dimensionless operation $q\to qM$. Thus we get using eqns. (20) and (22):
\begin{eqnarray}
p_t=g_{tt}\dot{t}+qA_{t}=-(1-\frac{2}{r}+\frac{Q^2}{r^2})\dot{t}-\frac{Qqr}{\Sigma}=-E, \nonumber\\p_{\phi}=g_{\phi \phi}\dot{\phi}+qA_{\phi}=r^2 sin^2\theta \dot{\phi}+\frac{qr^2 B_{0}}{2}=L,
\end{eqnarray}
where $L$ and $E$ are the angular momentum and constant specific energy of the system, respectively. Thus, the Hamiltonian, being identical to -$1/2$, is
\begin{eqnarray}
\mathbb{H}=\frac{1}{2}[\frac{r^2-2r+Q^2}{r^2}p^2_{r}+\frac{p^2_{\theta}}{r^2}-\frac{r^2E^2}{r^2-2r+Q^2}\nonumber\\
+\frac{1}{r^2\sin^2 \theta}(L-\frac{qr^2 B_{0}}{2})^2].
\end{eqnarray}
The above eqn. (24) is equivalent to
\begin{eqnarray}
\frac{(r^2-2r+Q^2)^2}{r^4}p^2_{r}+\frac{(r^2-2r+Q^2)}{r^4}p^2_{\theta}\nonumber\\=E^2-V^2_{\rm eff},\nonumber\\
V^2_{\rm eff}=\frac{r^2-2r+Q^2}{r^2}[1+\frac{1}{r^2}(L-\frac{qr^2 B_{0}}{2})^2],
\end{eqnarray}
where $V_{\rm eff}$ is the effective potential at the equatorial plane $\theta=\pi/2$.

We now study the innermost stable circular orbits at the equatorial plane, considering the values $L=3.4643, B_{0}=-10^{-3}, q=10^{-3}, Q=1$ (geometric units of length) \cite{qqq}. Then the feasible values of the radius of the circular orbit, for which it exists, i.e., $\dot{V}=0$, are $r=2.53579$ and $9.46556$. If $\ddot{V}>0$ then the orbit is stable. It is found that for only $r=9.46556$, $\ddot{V}=0.0008>0$ and the orbit is stable. It follows that for Kerr spacetime when $L=3.4643, Q=0, B_{0}=0$, we find the feasible value as $r=9.46336$ when $\ddot{V}=1.1\times 10^{-9}>0$ and a stable orbit exists. In the special case when $\ddot{V}=0$, the stable circular orbit is the innermost one. As an example, $r=1.5$ is found to be the radius of the innermost stable circular orbit (ISCO) of the system (24) with $B_{0}=0, Q=0, L=0$, i.e., of the Schwarzschild spacetime.

Despite the test motion being regular at the equatorial plane, there is a possibility of occurrence of chaos for generic orbits of test particles. This can occur with and without effects of rotation or electric charge, gravitational perturbations and imposed external electromagnetic fields [\cite{qqq},\cite{ttt}].

\section{Final Remarks}

~~~We have developed a feasible model of a compact star under $f(R,T)$ gravity in the presence of a magnetic fluid. These solutions are important for describing the interior of compact objects like neutron stars and quark stars. The magnetic field profile in the equatorial plane is also calculated. Various physical aspects of the compact star model under $f(R,T)$ gravity are elucidated.

We present suitable models of the neutron stars that are tested for some known compact
objects.  Stars of known geometry are analysed here as the equation of state is not known. 
The radii of the compact stars, namely, neutron stars, are also estimated here for known
mass $M$ with a given radius $R_1$ as per available data. By considering the observed masses of the compact objects, namely, X-ray pulsars Her X-1, 4U 1700-37 and neutron stars J1518+4904, J1748-2021B, we analyse the magnetic field of the star. We obtain a class of magnetised compact star models. The stellar models obtained here can accommodate highly  compact magnetised objects. 

Depending on the values of the constants $a_1$ and $b_1$ emerging from the $f(R,T)$ gravity, it is possible that there may be a significant effect on the interior solution of the magnetic field. This can enhance the expected significant difference between the average interior magnetic field strength and the surface magnetic field strength \cite{tt,uu}.

Future work can clarify the effect of this model on 
the magnetodipolar energy loss formula \cite{jj}, the mass-radius relationship of magnetised compact stars, the generation of magnetic fields in dense quark matter in HS/QS \cite{rr}, and the electromagnetic flashes of magnetars \cite{ss}.   

\section{Acknowledgments}
~~~SI is thankful to his wife Mrs. Tania Banu for her encouragement throughout and also to S. Datta, Md. A. Shaikh and Prof T. K. Das for providing some useful insights in the paper. We are grateful to the referees for their valuable comments.

\section{Conflict of Interest}
~~~Authors declare there is no conflict of interest.

\end{document}